\documentclass[journal,10pt]{IEEEtran}

\usepackage{amsmath}
\usepackage{amsthm}
\usepackage{amsfonts}
\usepackage{amssymb}
\usepackage{makeidx}
\usepackage{cite}
\usepackage{graphicx}
\usepackage{mathtools}
\usepackage{cases}
\usepackage{color}
\usepackage{bbm}
\usepackage[normalem]{ulem}
\usepackage{braket}
\usepackage{epstopdf}

\newtheorem{theorem}{Theorem}

\newtheorem{lemma}{Lemma}

\newtheorem{proposition}{Proposition}
\newtheorem{remark}{Remark}

\DeclareMathOperator{\cV}{\mathcal{V}}

\DeclareMathOperator{\cL}{\mathcal{L}}
\DeclareMathOperator{\cS}{\mathcal{S}}
\DeclareMathOperator{\cP}{\mathcal{P}}

\DeclareMathOperator{\bR}{\mathbb{R}}
\DeclareMathOperator{\bC}{\mathbf{C}}

\DeclareMathOperator{\bP}{\mathbf{P}}
\DeclareMathOperator{\ind}{\mathbbm{1}}
\DeclareMathOperator{\bE}{\mathbf{E}}
\DeclareMathOperator{\bZ}{\mathbb{Z}}
\newcommand*\diff{\mathop{}\!\mathrm{d}}

\newcommand*\nnb{\nonumber}

\newcommand{\ea}{\stackrel{(\text{a})}{=}}
\newcommand{\eb}{\stackrel{(\text{b})}{=}}
\newcommand{\ec}{\stackrel{(\text{c})}{=}}

\title{Poisson Cox Point Processes for Vehicular Networks}
\author{
	Chang-Sik~Choi~and~Fran{\c{c}}ois~Baccelli
\IEEEcompsocitemizethanks{\IEEEcompsocthanksitem 	{Chang-sik Choi is with the Department of ECE, The University of Texas at Austin, TX, USA (email: chang-sik.choi@utexas.edu). Fran{\c{c}}ois baccelli is with the Department of Mathematics and the Department of ECE, The University of Texas at Austin, TX, USA (email: baccelli@math.utexas.edu)} }
}
\begin{document}
	
	\maketitle
	\begin{abstract}
		This paper analyzes statistical properties of the Poisson line Cox point process useful in the modeling of vehicular networks. The point process is created by a two-stage construction: a Poisson line process to model road infrastructure and independent Poisson point processes, conditionally on the Poisson lines, to model vehicles on the roads. We derive basic properties of the point process, including the general quadratic position of the points, the nearest distance distribution, the Laplace functional, the densities of facets of the Cox-Voronoi tessellation, and the asymptotic behavior of the typical Voronoi cell under vehicular densification. These properties are closely linked to features that are important in vehicular networks.
	\end{abstract}

	\section{Introduction}
In vehicular networks, both vehicle distribution and road layout affect the communication performance. Spatial models representing the motion of vehicles were proposed in \cite{4382911,4581649}. Later, the locations of vehicles on \emph{one road} were modeled by a Poisson point processes (PPP) on the line in \cite{blaszczyszyn2009maximizing,blaszczyszyn2013stochastic}. There, the Laplace functional of PPP was leveraged to derive the signal-to-interference-plus-noise ratio (SINR) and the average throughput in such networks. In the Euclidean plane, a Poisson line Cox point process was introduced in \cite{baccelli1997stochastic}, where the vehicles are on \emph{multiple roads}, modeled by a Poisson line process (PLP). Subsequently, Poisson line Cox point processes were used to analyze vehicular networks in \cite{morlot2012population} and to analyze heterogeneous cellular networks in \cite{choi2017analytical}. In spite of its accurate representation of the coupled structure of vehicles and roads (Fig. 1), the Poisson line Cox point process was significantly less utilized compared to the planar {PPP}, partly because only a few analytical results are available. 

\par The main aim of the present paper is to provide statistical and analytical properties of the Poisson line Cox point process model that are useful for the study of vehicular networks where vehicles are transmitters. We present several mathematical results on the Poisson line Cox point process, including the nearest distance distribution, the Laplace functional (under the stationary measure and under the Palm), the densities of facets in the Cox-Voronoi tessellation, and the asymptotic shape of the typical cell under vehicle densification. The results provided in this paper are important to analyze vehicular networks in general. For instance, the Cox-Voronoi tessellation characterizes the association regions of vehicular transmitters and therefore their facet densities allow one to assess the amount of information exchanged over the association region. Furthermore, similar to the work \cite{blaszczyszyn2009maximizing} where results on the Laplace functional of PPP were used to derive the distribution of the Shannon rate of the typical link, the presented results on the nearest distance and the Laplace functional will be useful to characterize the performance of vehicular networks, in particular, to derive the distribution of interference and the SINR distribution of the typical user\cite{choi2017analytical}. 

\begin{figure}
	\centering
	\includegraphics[width=1\linewidth]{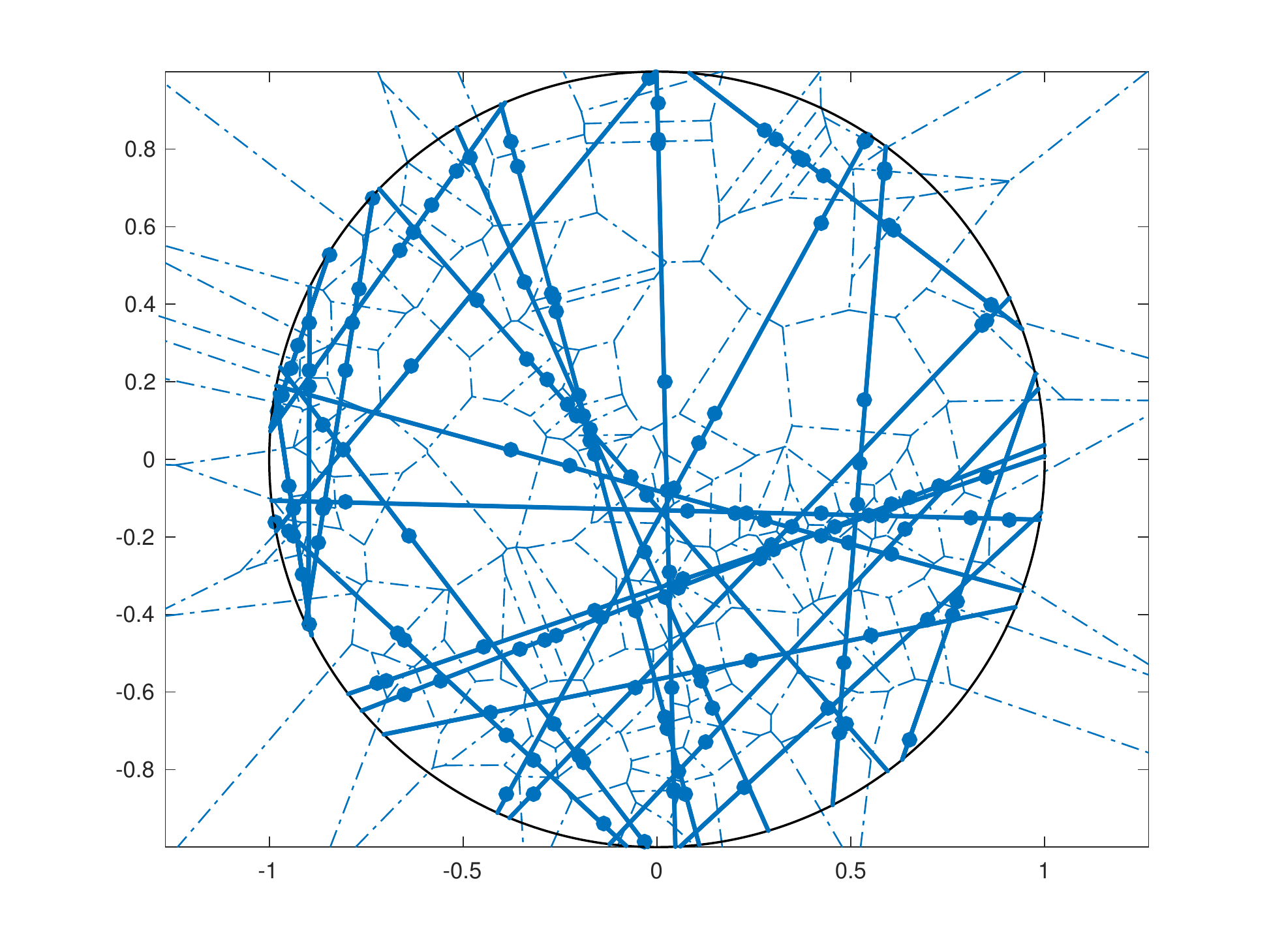}
	\includegraphics[width=1\linewidth]{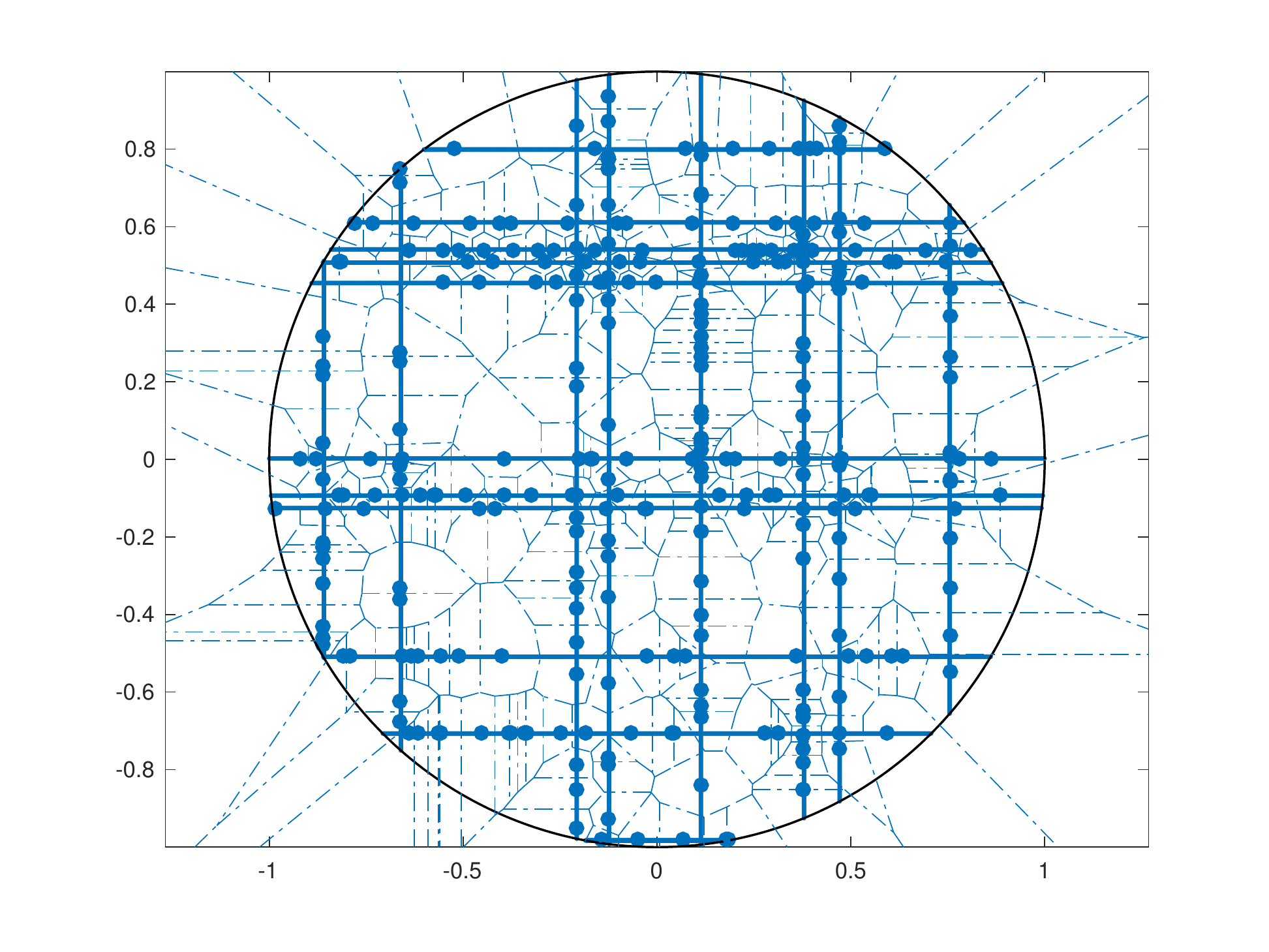}
	\caption{Illustration of roads modeled by PLP $ \Psi $ (solid lines), vehicles modeled by Poisson line Cox point process $ \Phi $ (dots) and association regions modeled by Voronoi tessellation with respect to (w.r.t.) $ \Phi $}%
	\label{fig:coxvoronoi}
\end{figure}
	\section{System Model for Vehicular Networks}
To model roads, we consider a PPP $ \Xi $ on the cylinder set $ \mathbf{C}:=\bR \times [0,\pi) $ with intensity measure 
$ \Lambda_{\Xi}(\diff r \diff \theta)=\lambda_l\diff r G(\diff \theta). $
Each point $ {(r_i,\theta_i)}_{i\in\bZ}$ of $ \Xi $ corresponds to a line 
where the parameters $ r_i $ and $ \theta_i $ are the distance from the origin to the line and the angle between the line and the $ x-$axis measured in a counterclockwise direction, respectively. The collection of lines in the Euclidean plane is referred to as a PLP $ \Psi $ and this line process is stationary \cite{chiu2013stochastic}. Furthermore, if $ G(\diff \theta)={\diff\theta}/{\pi}, $ the PLP is isotropic (Fig. 1 top); if $ G(\diff\theta)= 0.5\delta_{0}+0.5\delta_{\pi/2},$ the lines are horizontal and vertical (Fig. 1 bottom), which is referred to as the Manhattan PLP.  This paper considers $ G(\diff\theta)=\diff\theta/\pi $only.  However, the techniques can be easily extended to the Manhattan case.
\par Conditionally on the lines of $ \Psi $, independent stationary PPPs with intensity $ \mu $, denoted by $ \{\phi\}_{} $, are created on the lines to model vehicles on roads. The distance between two consecutive points on the same line hence follows the exponential distribution with parameter $ \mu $. The PPPs on different lines are assumed to be conditionally independent. The collection of points is referred to as the Poisson line Cox point process $ \Phi. $ Fig. 1 illustrates a realization of $ \Phi $ and its Voronoi tessellation.

%We will consider only $ \bR^2 $ in the rest of the paper.
 %For a point process $ \Xi $ and its Voronoi tessellation, if $ \Xi $ is in general position then, the number of edges from a vertex of the Voronoi tessellation is three  with probability one. We investigate this property of the Voronoi tessellation w.r.t. $ \Phi $. 

%In addition, we will consider the following object induced by the tessellation. 

%The following lemma is classical. We give its proof in order to better explain the proof of later results.
\section{Result}
\subsection{Stationarity, General Quadratic Position, and Facets}
	 \begin{lemma}\label{Lemma:1}\textbf{Stationarity}
	 The distribution of $\Phi$ is translation and rotation invariant. 
	%Similarly, the intensity of the proposed Cox process $ \Phi_3 \in\bR^2$ is given by $ \gamma\lambda_l. $  
\end{lemma}
See Appendix \ref{A:1} for the proof.  
%\begin{IEEEproof}
% Appendix A.
%\end{IEEEproof}
\begin{lemma}
	\textbf{Density} The density of $ \Phi $ is $ \mu\lambda_l. $  
\end{lemma}
See Appendix \ref{A:2} for the proof.
%\begin{IEEEproof}
%
%\end{IEEEproof}
%\begin{definition}\label{d2}
%	The points of a simple point process are in weak general quadratic position if 
%	there exists no 4 points lie on a circle.  
%\end{definition}
\begin{lemma}\label{L:GQP}
	\textbf{General quadratic position} With probability one, no four points of $ \Phi $ lie on a circle. 
\end{lemma}
See Appendix \ref{A:3} for the proof. 
\par Recall the Voronoi cell of $ X_i $ w.r.t. $ \Phi $ is 
$ \mathcal{V}(X_i)=\{x\in\bR^2/ \|x-X_i\|\leq \inf_{Y_i\in\Phi\setminus X_i}\|x-Y_i\|\}, $
whereas the Voronoi tessellation w.r.t. $ \Phi $ is the collection of the Voronoi cells, $ \cV=\cup_{X_i\in\Phi} \cV({X_i}) $\cite{chiu2013stochastic}. Fig 1. illustrates the Cox-Voronoi tessellation. The tessellation is comprised of random objects such as vertices, edges, and 2-dimensional facets. They are referred to as the $ (0,1,2) $-facets of the tessellation and are point processes in the space of random closed set \cite{moller2012lectures}. 

\begin{proposition}\textbf{Density of vertices, edges, and faces}
	The densities of the (0,1,2)-facets of the Cox-Voronoi tessellation are 
$ 	\lambda^{(0)}=2\mu \lambda_l,\,
	\lambda^{(1)}=3\mu\lambda_l,\,
	\lambda^{(2)}=\mu \lambda_l,
 $
	respectively. 
\end{proposition}
See Appendix D for the proof. 
\begin{remark}
	The PPP was extensively utilized to model cellular networks. Its Voronoi tessellation represents the coverage regions of base stations when each user is associated with the closest base station. Similarly, the Cox-Voronoi tessellation in this paper characterizes the association region of vehicular transmitters. Note that the  Cox-Voronoi tessellation is qualitatively very different from the Poisson-Voronoi tessellation created by a planar PPP. For more on this qualitative difference, See Section \ref{S:SC}.
\end{remark}
 In the following, we analyze the nearest distance distribution and the Laplace functional under the stationary and under the Palm distribution. The analysis is based on Palm calculus: under the Palm distribution of $ \Phi $, one finds the \emph{typical vehicle} at the origin and the \emph{typical road} containing the origin. 
\subsection{Nearest Distance Distribution and Laplace Functional}

 %The point on line $ i $ nearest to the origin is denoted by $ \tilde{X_{i,0}} $; its left and right points are denoted by $ X_{i,0},X_{i,-1},X_{i,-2},\ldots $ and $ X_{i,1},X_{i,2},X_{i,3},\ldots $, respectively.
%Figure \ref{fig:plporientation} illustrates $ \{X_{i,j}\}_{i,j\in\bZ} $. 

\begin{figure}
	\centering
	\includegraphics[width=1\linewidth]{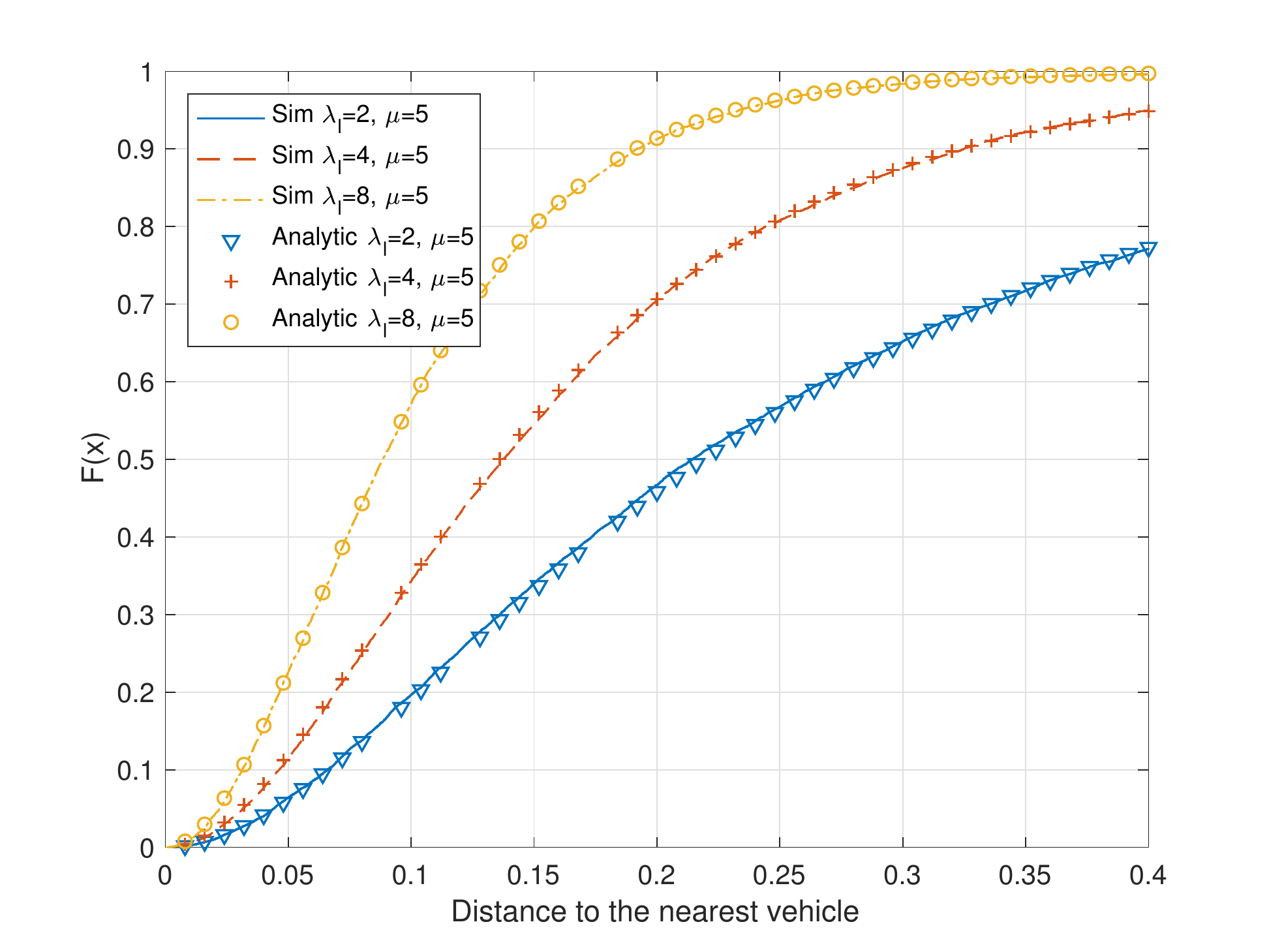}
	\caption{Distance distribution from a typical point to its nearest vehicle}
	\label{fig:nearest_noPalm}
\end{figure}
\begin{figure}
	\centering
	\includegraphics[width=1\linewidth]{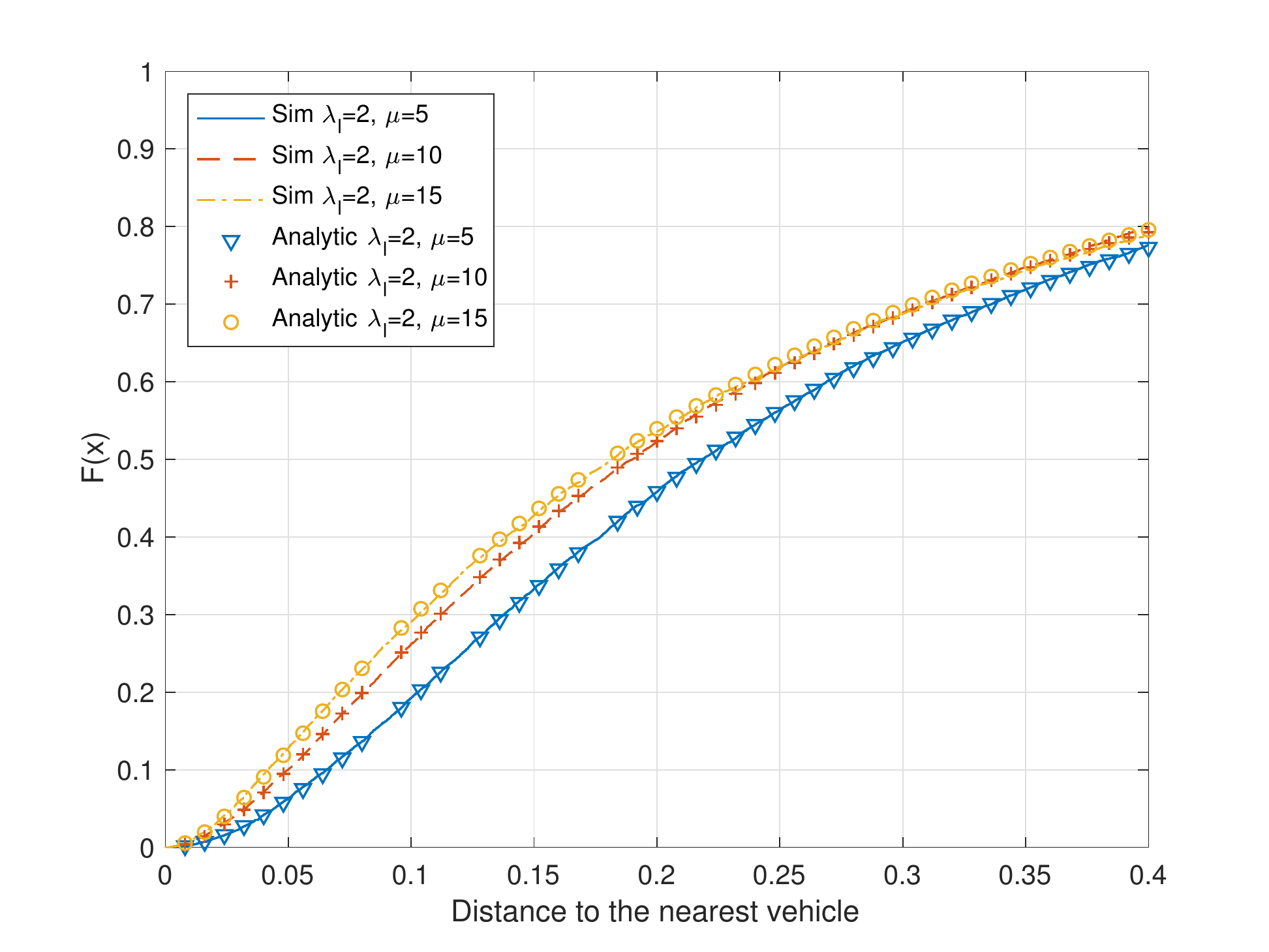}
	\caption{Distance distribution from a typical point to its nearest vehicle}
	\label{fig:nearest_noPalm2}
\end{figure}

\begin{figure}
	\centering
	\includegraphics[width=1\linewidth]{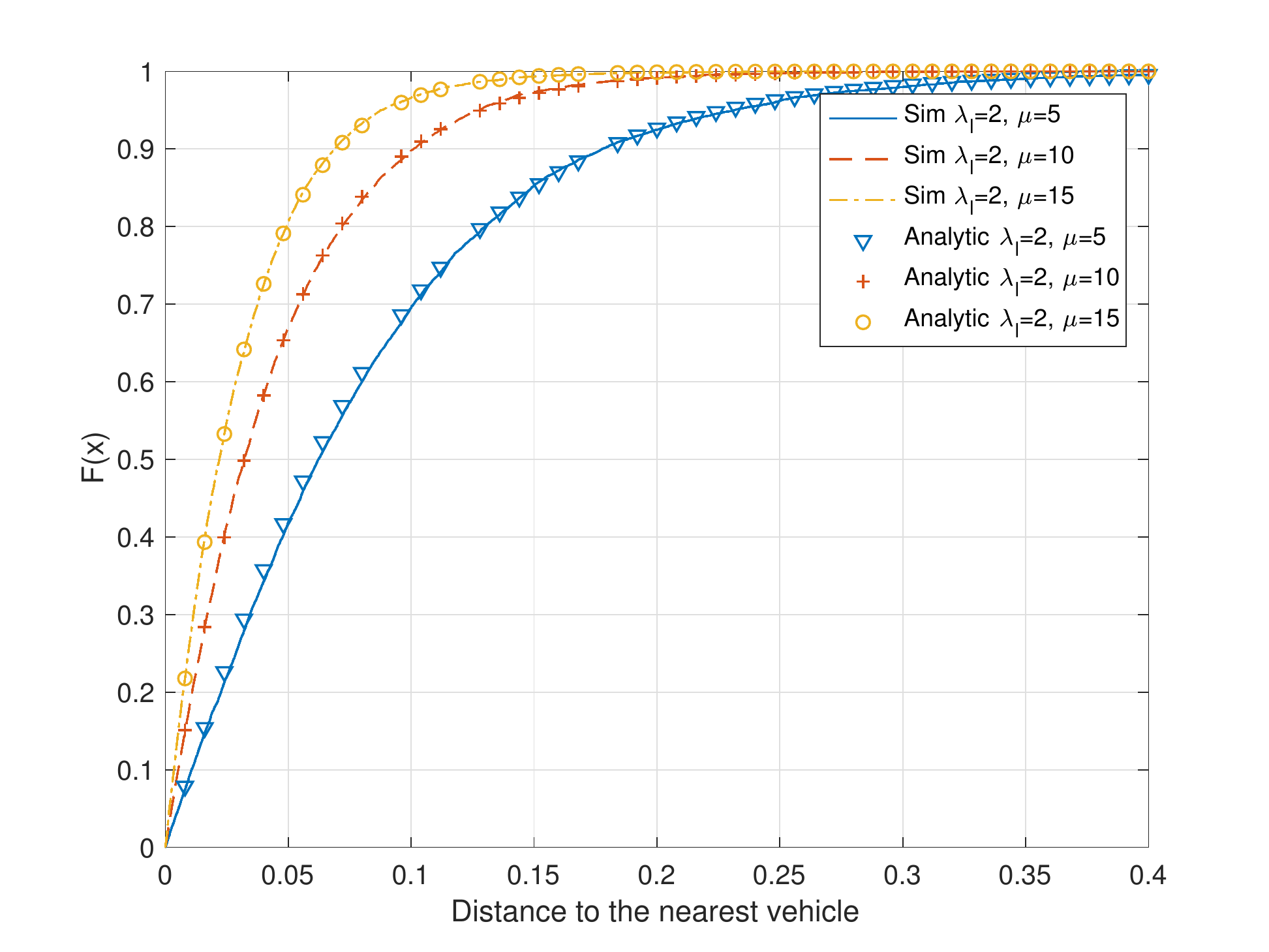}
	\caption{Distance distribution from a typical vehicle to its nearest vehicle}
	\label{fig:nearest_Palm}
\end{figure}
\begin{figure}
	\centering
	\includegraphics[width=1\linewidth]{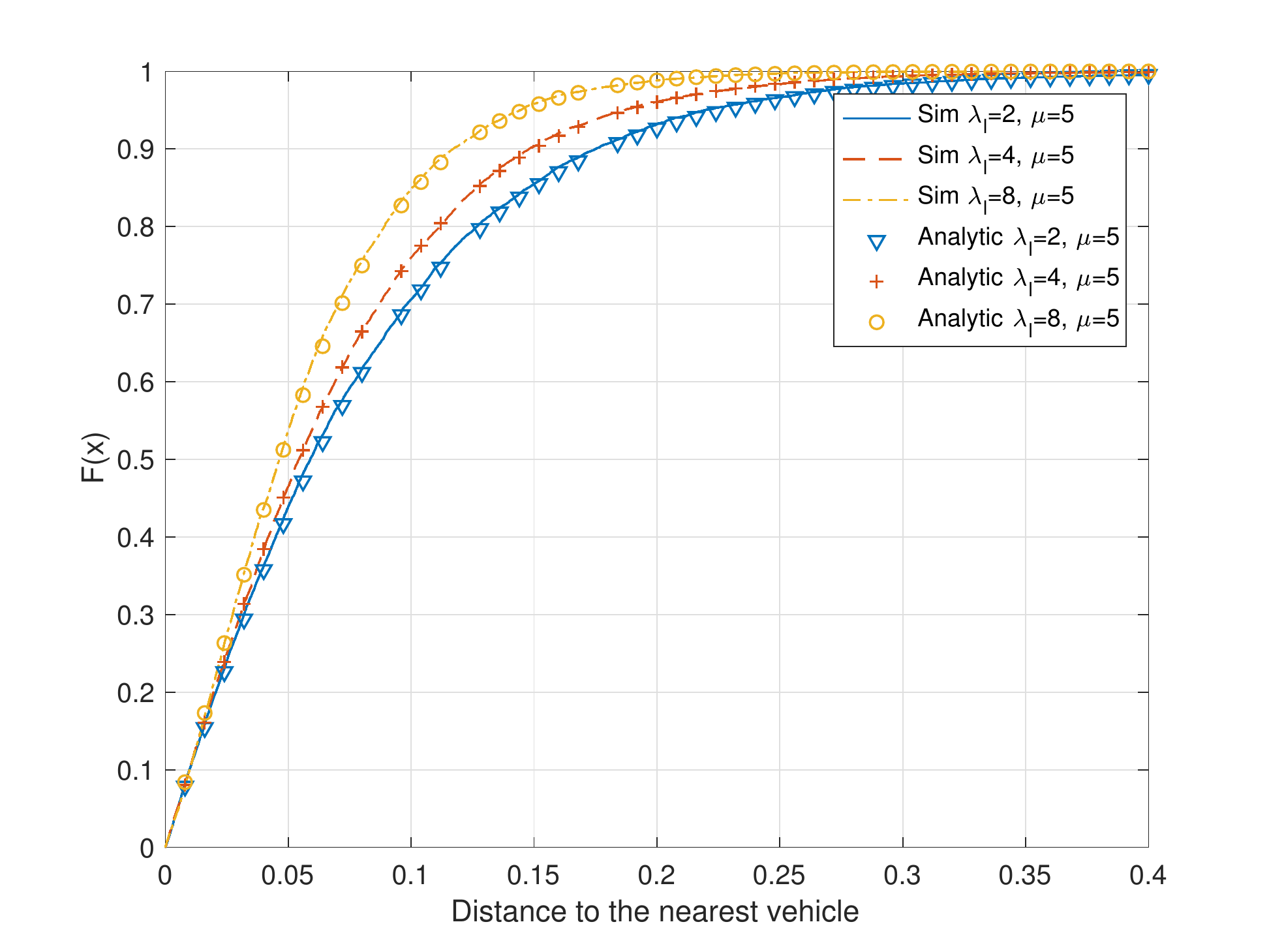}
	\caption{Distance distribution from a typical vehicle to its nearest vehicle}
	\label{fig:nearest_Palm2}
\end{figure}

\begin{lemma}\textbf{Nearest distance distribution}
		The distribution of the distance from an arbitrary point in the Euclidean plane to the nearest point of $\Phi $ is given by
	\begin{equation}
		\bP(R<r)=1-e^{-2\lambda_l\int_{0}^{r}1-e^{-2\mu\sqrt{r^2-{u}^2}}\diff u}.\label{2-2}
	\end{equation}
	Under the Palm distribution, it is given by 
 \begin{equation}\label{eq.2}
 \bP(R<r)=1-e^{-2\mu r-2\lambda_l\int_{0}^{r}1-e^{-2\mu\sqrt{r^2-{u}^2}}\diff u}.  
 \end{equation}
\end{lemma}
\begin{IEEEproof}
Using the stationarity of $  \Phi $, we have 
	\begin{align*}
	\bP(R\geq r)&\stackrel{(a)}{=}\bE_{\Phi}\left[\prod_{X_k\in\Phi}\mathbbm{1}_{\|X_k\|>r}\right]\\
	&\stackrel{(b)}{=}\bE_{\Psi}\left[\prod_{i\in\bZ}\bE_{\phi_i}\left[\left.\prod_{j\in\bZ}\mathbbm{1}_{\|X_{i,j}\|>r}\right|\Psi\right]\right]\\
	&\stackrel{(c)}{=}\bE_{\Psi}\left[\prod_{i\in\bZ}\bP\left(\|X_{i,0}\|\wedge \|X_{i,1}\|\geq \sqrt{r^2-r_i^2}\right)\right]\\
	&\stackrel{(d)}{=}\bE_{\Psi}\left[\prod_{i\in\bZ}\exp\left(-2 \mu \sqrt{r^2-|r_i|^2}\right)\mathbbm{1}_{-r<r_i<r}\right]\\
	&=\exp\left(-2\lambda_l\int_{0}^{r}1-e^{-2\mu\sqrt{r^2-{u}^2}}\diff u\right),
	\end{align*}
where $ x\wedge y $ denotes the minimum of $ x $ and $ y. $
We have (a) from the fact that all points of $ \Phi $ are at distance more than $ r $ and (b) by conditioning on $ \Psi$ and writing point $ X_{k} $ as $ X_{i,j}, $ where index $ i $ means the line on which $ X_{k} $ is located. We have (c) when denoting by $ X_{i,0} $ and $ X_{i,1} $ the two nearest points of $ \Phi $ on line $ i $ that are on the opposite sides around the closest point  from the origin on line $ i $. Finally, (d) follows from the fact that the minimum of two exponential random variables with parameter $ \mu $ is an exponential with parameter $ 2\mu $. Applying the Laplace transform of the PPP \cite{chiu2013stochastic} gives the result. 
\par Note that, under the Palm distribution, the nearest distance is given by minimum of (A) the distance from the typical vehicle to the nearest point on the typical line and (B) the distance given in Eq. \eqref{2-2}. Since (A) follows an exponential distribution with parameter $ 2\mu,  $ the nearest distance under the Palm is obtained. 
\end{IEEEproof}
The exactness of the derived lemma is illustrated in Figures \ref{fig:nearest_noPalm}-\ref{fig:nearest_Palm2}. To empirically verify the derived analytical result, large-scale system simulation is designed to empirically evaluate the distance from the origin to the nearest point. These figures clearly demonstrate that the analytical formulas are correct and accurate. To minimize edge-effect in a finite simulator, simulations are designed in a disk of radius $ 5 $. Interestingly, as Figure \ref{fig:nearest_noPalm2} shows, an increase of the linear density of vehicles, namely  $ \mu, $ has a diminishing impact on the distance distribution function. This occurs because the distance to the vehicle is strictly lower bounded by the distance to the road. For instance, even when $ \mu $ is infinity, the distance  to the nearest vehicle is given by the distance to the nearest line, which is distributed according to an exponential random variable with parameter $ 2\lambda_l. $
\begin{remark}
	Under the Palm probability, there exists a typical vehicle and a typical line at the origin. Therefore, the distance from the origin to a vehicle is interpreted into the distance from a typical vehicle to its nearest vehicle. Figures \ref{fig:nearest_Palm} and \ref{fig:nearest_Palm2} illustrate the distance from a typical vehicle to the nearest vehicle. By comparing Figures \ref{fig:nearest_Palm} and \ref{fig:nearest_Palm2} to their counterparts,  Figures \ref{fig:nearest_noPalm} and \ref{fig:nearest_noPalm2}, it is very clear that under the Palm distribution, higher distribution functions are acquired. It happens because the existence of a vehicle at the origin allows a typical road and the points on the typical road increase the chance of having a shorter distance. In other words, the distance from the typial vehicle stochastically dominates the distance from the origin. 
\end{remark}

%\begin{figure}
%	\centering
%	\includegraphics[width=1\linewidth]{nearest}
%	\caption{}
%	\label{fig:nearest}
%\end{figure}

\begin{lemma}\textbf{Laplace functional}
	Consider a function $ f:\bR^2\times \to [0,1]$. The Laplace functional of the Poisson line Cox point process is given by Eq. \eqref{Laplace}. If $ f $ is radially symmetric, then the Laplace functional is given by Eq. \eqref{Laplace2}.  {Under the Palm distribution, it is given by Eqs. \eqref{add}, and \eqref{add2}, respectively.}
	
\end{lemma}
\begin{figure*}
	\begin{align}
	\cL(f)&=\exp\left(\!-\frac{2\lambda_l}{\pi}\int_{0}^\infty\int_{0}^\pi 1-\exp\left(\!-2\mu\int_{0}^\infty 1-e^{- f(t \cos(\theta)-r\sin(\theta), t\sin(\theta)+r\cos(\theta))}\diff t\right)\diff \theta \diff r\right).\label{Laplace}\\
		\cL(\tilde{f})&=\exp\left(\!-{2\lambda_l}\int_{0}^\infty 1-\exp\left(-2\mu\int_{0}^\infty 1-e^{-\tilde{f}(\sqrt{t^2+r^2})}\diff t\right)\diff r \right).\label{Laplace2}
	\end{align}
	\vspace{-1em}
\rule{\textwidth}{0.2pt}	\vspace{-1em}
\end{figure*}

\begin{IEEEproof}
	The Laplace functional is given by 
	\begin{align*}
	\cL(f)=&\bE_{\Phi}\left[e^{-\sum_{X_{i,j}\in\Phi}f(X_{i,j})}\right]\\
	%&=\bE_{\Phi}\left[\prod_{X_{i,j}\in\Phi} e^{-sf(X_{i,j})}\right]\\
	&\stackrel{(a)}{=}\bE_{\Psi}\left[\prod_{i\in \bZ}\bE_{\phi_i}\left[\left.\prod_{j\in\bZ}e^{-f(X_{i,j})}\right|\Psi\right]\right]\\
	&\stackrel{(b)}{=}\bE_{\Psi}\!\left[\prod_{i\in\bZ}\exp\left(-\mu\int_{\bR}\left(1-e^{-\tilde{g}(i,t)}\right)\diff t\right)\right]\\
	&\stackrel{(c)}{=}\exp\left(-\frac{\lambda_l}{\pi}\int_{\bR}\int_{0}^\pi 1\!-e^{\!-{\mu}\!\int_{\bR}\!1-e^{-\tilde{h}(r,\theta,t)}\diff t}\!\diff \theta \diff r\right)\!,
	\end{align*}
	where 
	\begin{align}
		&\tilde{g}(i,t)=f(t\cos(\theta_i)-|r_i|\sin(\theta_{i}), t\sin(\theta_i)+|r_i|\cos(\theta_i))\nnb,\\
		&\tilde{h}(r,\theta,t)=f(t\cos(\theta)-|r|\sin(\theta), t\sin(\theta)+|r|\cos(\theta)).\nnb
	\end{align}To obtain (a), we condition with respect to the PLP; to get (b) and (c), the Laplace functionals of PPPs are used. 
	 \par 
	If $ f(\cdot) $ is radially symmetric,  $ f(x,y)\equiv\tilde{f}(\sqrt{x^2+y^2}) $, the Laplace functional is given by 
	\begin{align*}
	\cL(\tilde{f})%&=\bE_{\Psi}\left[\prod_{i\in \bZ}\bE_\phi\left[\left.\prod_{j\in\bZ}e^{-\tilde{f}(\sqrt{r_i^2+t_j^2})}\right|\Psi=\{l_i\}_{i\in\bZ}\right]\right]\\
	&=\bE_{\Psi}\left[\prod_{i\in \bZ}\exp\left(-\mu\int_{\bR}1-e^{-\tilde{f}(r_i^2+t^2)}\diff t\right)\right]\\
	&=\exp\left(-\frac{\lambda_l}{\pi}\int_{\bR}1-e^{-{\mu}\int_{\bR}1-e^{-\tilde{f}(r^2+t^2)}\diff t}\diff r \right).
	\end{align*}
	\par Under the Palm distribution, the Laplace functional $ \hat{\cL}(f) $ is
	\begin{align}
			\hat{{\cL}}(f)&=\bE_{\Phi}\left[e^{-\sum_{X_{i,j}\in\Phi}f(X_{i,j})-\sum_{X_{0,j}\in\phi_0}f(X_{0,j})}\right]\nnb\\
			&=\cL(f)\bE\left[e^{\sum\limits_{X_{0,j}\in\phi_0}f(X_{0,j})}\right]\nnb\\
			&=\cL(f)\int_{0}^{\pi}\pi^{-1}e^{-\mu\int_{\bR}1-e^{-f(t\cos(\theta),t\sin(\theta))}\diff t}\diff \theta,\label{add}\\
			&\stackrel{(a)}{=}\!\cL(f)e^{-\mu\int_{\bR}1-e^{-\tilde{f}(t)}\diff t}\label{add2}
	\end{align}
	because the points on the typical line containing the origin are independent of $ \Phi $. We have $ (a) $ only if $ f(\cdot) $ is radially symmetric. 
%Therefore, the Palm versions of the Laplace functionals are given by Eq. \eqref{Laplace} multiplied by Eq. \eqref{add} and Eq. \eqref{Laplace2} multiplied by Eq. \eqref{add2}, respectively.
\end{IEEEproof}
{The Laplace functional and its Palm version allow us to analyze vehicular networks under two different perspectives as explained in the following remark.}
\begin{remark}
	If $ f(x,y)=(x^2+y^2)^{-\frac{\alpha}{2}} $, the Laplace functional in Eq. \eqref{Laplace2} becomes the Laplace of the total interference seen by an arbitrary point of the Euclidean plane, when the received signal power attenuates according to the distance-based path loss function $ f(x,y) $. \par{Furthermore, if one wants to capture the interference seen by an arbitrary vehicle, the Palm version of the Laplace functional yields the desired result.} 
\end{remark}
%\begin{corollary}
%	Consider a marked Poisson point process $ \Phi_u $ with intensity $ \lambda_u$ and a functional on the point process $ X_i\in\Phi_u (\mathcal{V}(\Phi))$ denoted by $ G(X) $.  The expectation of the functional with respect to the Cox process is given by
%	\begin{equation}
%		\bE_{\Phi_u}\left[\sum_{X_i\in \Phi}\mathbbm{1}_{Z_i\in\mathcal{V}(0,\Phi)}g(Z_i)\right]=\bE\left[\int \mathbbm{1}_{}\right].
%	\end{equation}
%\end{corollary}
%\begin{IEEEproof}
%%	The additive functional inside the Voronoi is given by
%%	\begin{align*}
%%		\bE\left[\right]	
%%	\end{align*}
%\end{IEEEproof}
%\subsection{Voronoi tessellation of the Cox point process}

\subsection{Asymptotic Shape of the Typical Voronoi Cell}\label{S:SC}
The typical Voronoi cell is the Voronoi cell of the origin under the Palm distribution of $ \Phi $, \cite{baccelli2009stochastic}.  Under the Palm distribution, the typical point is located at the origin and the typical line contains the origin. Below, we provide a new numbering of the points of $ \Phi $ \emph{seen} from the origin. Since $ \Phi $ is rotation invariant, we can assume, without loss of generality, that the typical line is the $ x$-axis. Then, note that the $ x$-axis dissects the Euclidean space into an upper half plane $ H_+ $  and a lower half plane $ H_-. $ Let us imagine a ball in $ H_+ $ tangent to the $ x$-axis at the origin and grow its radius. The first point at which it meets another line is denoted by $ \tilde{X}_{1,0} $. Similarly, imagine a ball in $ H_- $ tangent to the $ x$-axis at the origin and grow its radius. The first point at which it meets another line is denoted by $ \tilde{X}_{-1,0} $. This method defines all points $ \tilde{X}_{i,0} $ for all $ i\in\bZ. $ The line containing $ \tilde{X}_{i,0}  $ is denoted by $ l_{i} $. Subsequently, for the points on each line $ l_{i}, $ we again use the well-known convention: around $ \tilde{X}_{i,0}, $ the points on its right are denoted by $ X_{i,1},X_{i,2},X_{i,3},\ldots $ and the points on its left are denoted by $ X_{i,0},X_{i,-1},X_{i,-2}\ldots $.\footnote{The above numbering is handy to characterize the typical cell but does not produce a unique numbering; note that each Cox point has two sets of indexes since every line is eventually met twice by balls in $ H_+ $ and $ H_- $, respectively.}
%\begin{figure}
%	\centering
%	\includegraphics[width=0.9\linewidth]{Figure/Voronoinumbering}
%	\caption{The points $ \{\tilde{X}_{k,j}\}_{k\neq i} $ indicate the nearest points of lines from $ X_{i,j} $. }
%	\label{fig:voronoinumbering}
%\end{figure}

%\par Then, conditionally on the $ \Psi, $ the intensity parameter of $ \phi $ goes to infinity in order to obtain the shape of the Voronoi cell at points in $ \Phi. $

%The first index is increasing with respect to radius from point $ X_{i,j} $ where the circles are on the line $ i $. The point of the lines that touches the circles are denoted by $ \{\tilde{X}_{k,j}\}_{k\neq i} $. With probability one, $ \tilde{X}_{k,j} $ are not points of $ \Phi. $ Then, we have the following lemma. %The other points %The asymptotic regime is of high importance in order to understand the coverage of wireless networks where the communication devices are placed along roads. When the transmitters are placed per a specific way, the coverage area of each transmitter corresponds to the proposed Cox Voronoi. 
	\begin{theorem}\label{theorem:2}\textbf{Convergence of the typical cell}
	As $ \mu\to\infty $, the typical Cox-Voronoi cell converges to a segment almost surely (a.s.) in the sense of the Fell topology. The segment is contained in the $ y$-axis. Its positive and negative parts follow independent exponential distributions with parameter $ 2\lambda_l $.% and the distance between the long edges follows independent Erlang distribution with parameter $ (2,2\mu) $.
\end{theorem}
\begin{IEEEproof}  
	 In order to show the covergence in the space of random closed set, we use the Painlev\'{e}-Kuratowski convergence \cite[Def. 5.5.1]{baccelli2009stochastic}. The typical  cell $ \cV(0) $ is defined by 
\begin{align}
\cV(0)&=\bigcap_{k,l\in\bZ^2\setminus \{0,0\}}\{y\in\bR^2/ \|y\|\leq \|y-X_{k,l}\|\},\nnb\\
&=\bigcap_{k,l\in\bZ^2\setminus \{0,0\}}H_{k,l}(\mu)\nnb\\
%&=\bigcap_{l\in\bZ\setminus \{0\}}H_{0,l}\bigcap_{k\in\bZ\setminus\{0\},l\in\bZ}H_{k,l}\nnb\\
&=\bigcap_{l\geq +1}H_{0,l}(\mu)\bigcap_{l\leq -1}H_{0,l}(\mu)\bigcap_{k\in\bZ\setminus\{0\},l\in\bZ}H_{k,l}(\mu)\nnb\\
&=H_{0,-1}(\mu)\bigcap H_{0,1}(\mu)\bigcap_{k\in\bZ\setminus\{0\},l\in\bZ}H_{k,l}(\mu)\nnb\\
&=\mathcal{P}_{0}(\mu)\bigcap_{k\in\bZ\setminus\{0\},l\in\bZ}H_{k,l}(\mu),\label{6}
\end{align}
where $ H_{k,l}(\mu) $ denotes the half plane associated with $ X_{k,l} $. To obtain Eq.\eqref{6}, we use 
\begin{equation}
\cap_{l\geq 1}H_{0,l} (\mu)= H_{0,1}(\mu) \text{ and } \cap_{l\leq -1}H_{0,l}(\mu) = H_{0,-1}(\mu)\nnb.
\end{equation}
Below, we use a coupling where the PPP for $ \mu'>\mu $ is obtained by adding an independent PPP of intensity $ \mu'-\mu $ to the PPP of intensity $ \mu. $
Since $ X_{0,1} $ and $ X_{0,-1} $ tend monotonically to the origin a.s.  as $ \mu\to\infty $, we have  $ \cP_{0}(\mu)\downarrow \mathcal{Y}$, the $ y$-axis where we denote by $ A_n\downarrow A $, the fact that sequence of sets $ A_n $ decreases to the set $ A $. Hence, $ \lim_{\mu\to\infty}\cP_{0}(\mu)=\mathcal{Y}$ in the Fell sense \cite[Cor. 3]{matheron1975random}. Moreover, since $ X_{1,0} $ and $ X_{1,1} $ tend monotonically to $ \tilde{X}_{1,0} $ a.s.,  we have 
\begin{align}
	\{ \mathcal{Y}\cap H_{1,0}(\mu) \}_\mu &\downarrow \mathcal{Y}\cap \tilde{H}_{1,0},\label{1}\\
	\{\mathcal{Y} \cap H_{1,1}(\mu) \}_\mu &\downarrow \mathcal{Y}\cap \tilde{H}_{1,0}\label{2},
\end{align}
where $ \tilde{H}_{1,0} $ is the half plane generated by $ \tilde{X}_{1,0} $.
\begin{figure}
	\centering
	\includegraphics[width=0.7\linewidth]{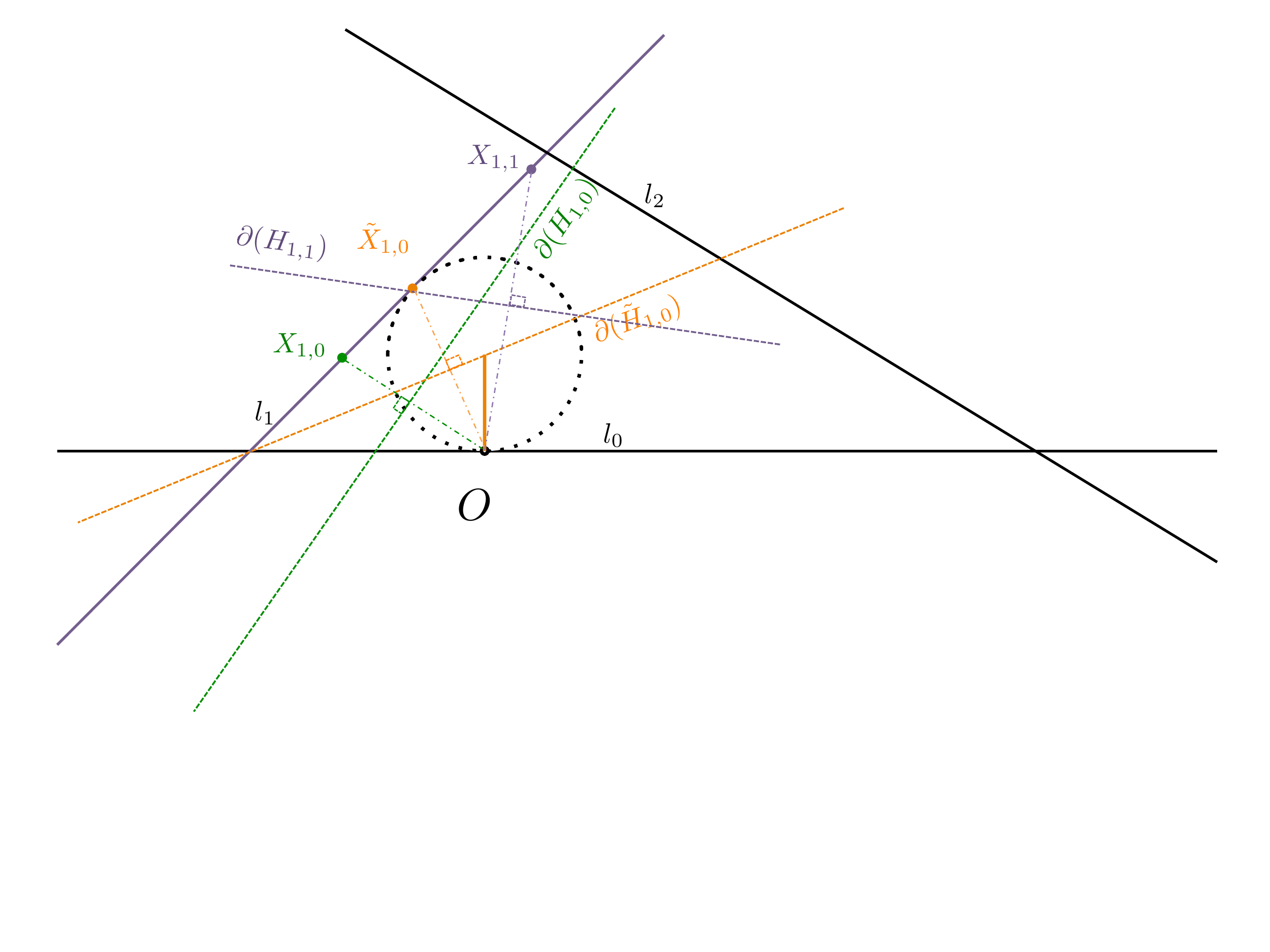}
	\caption{Half planes associated with two Cox points and the nearest points to the origin on line $ l_1 $, respectively.}
	\label{fig:prooffigure}
\end{figure}
Figure \ref{fig:prooffigure} illustrates the sets in question. More generally, 
 $\forall k\geq 1 $
\begin{align}
\{ \mathcal{Y}\cap H_{k,0}(\mu) \}_\mu &\downarrow \mathcal{Y}\cap \tilde{H}_{k,0},\label{3}\\
\{\mathcal{Y} \cap H_{k,1}(\mu) \}_\mu &\downarrow \mathcal{Y}\cap \tilde{H}_{k,0}.\label{4}
\end{align}
On the other hand, for all $ k\geq 1, $ the following relations hold 
\begin{align}
%	&\mathcal{Y} \cap \tilde{H}_{1,0}\subseteq \mathcal{Y} \cap_{l\neq 0,1} H_{1,l},\label{5}\\
	&\mathcal{Y} \cap \tilde{H}_{1,0}\subseteq \mathcal{Y} \cap_{k\geq 1} \tilde{H}_{k,0},\label{6'}\\
	&\mathcal{Y} \cap \tilde{H}_{k,0}\subseteq \mathcal{Y} \cap_{l\neq 0,1} {H}_{k,l}\label{6''}.
\end{align}
Using $ A_n\downarrow A $ and $ B_n\downarrow B $, $ \lim(A_n\cap B_n)=A\cap B $\cite[Cor. 3]{matheron1975random} and Eqs. \eqref{1}-\eqref{6''}, we have the following a.s. convergence
\begin{equation}
 \lim_{\mu\to\infty}\cP_0(\mu)\bigcap_{k\geq 1, l\in\bZ} H_{k,l}=\mathcal{Y}\cap \tilde{H}_{1,0}. \label{13}
\end{equation}
By similar arguments, we have %Moreover, since $ A_n\cap B_n \cap C_n= (A_n\cap B_n )\cap(A_n\cap C_n )$, we have 
\begin{equation}
 \lim_{\mu\to\infty}\cP_0(\mu)\bigcap_{k\leq -1, l\in\bZ} H_{k,l}=\mathcal{Y}\cap \tilde{H}_{-1,0}. \label{14}
\end{equation}
Hence, combining Eq. \eqref{13} and \eqref{14} gives %concludes that 
\begin{align}
	\lim_{\mu\to\infty}\cV(0)=\mathcal{Y}\cap \tilde{H}_{1,0}\cap \tilde{H}_{-1,0}=\mathcal{S}.
\end{align}
As a result, $ \cV(0) $ a.s.  converges to a one-dimensional {segment} in the Fell topology,  given by the intersections of a line $ \mathcal{Y}$ and two half planes, $ \tilde{H}_{0,-1}$ and  $\tilde{H}_{0,1}$, respectively. Moreover, $ |\cS|=|\cS_+|+|\cS_-|=|\mathcal{Y}\cap \tilde{H}_{1,0}\cap H_+|+ |\mathcal{Y}\cap \tilde{H}_{-1,0}\cap H_-|$. 
\begin{align}
\bP(|\cS_+|=l)&=\bP(\text{interior of }\Psi\cap B_{(0,l)}(l) \text{ is empty })\nnb\\
&=\bP(\Psi(C_l)=0)\nnb\\
&=1-\exp\left(-\frac{\lambda_l}{\pi} \int_{0}^\pi\int_{l(\cos(2\theta)-1)}^{l(\cos(2\theta)+1)}\diff r\diff \theta\right)\nnb\\
&=1-\exp(-2\lambda_l l),\nnb
\end{align}
where we use the fact that the ball centered at $ (0,l) $ with radius $ l $, $ B_{(0,l)}(l) $,  corresponds to  $ C_l:=l(\cos(2\theta)-1)\leq r\leq l(\cos(2\theta)+1)$ for $ \theta\in(0,\pi) $. Therefore, $ S_+ $ follows an exponential with parameter $ 2\lambda_l $. Similarly, $ S_- $ follows an independent exponential with parameter $ 2\lambda_l $. Consequently, $ |\cS|$ follows an Erlang distribution with parameters $ (2,2\lambda_l). $ 
\end{IEEEproof}
\begin{remark}
	The above quantification of the Cox-Voronoi cell in the asymptotic regime is interesting since the densification of transmitters is a key enabler for 5G cellular networks. Contrary to the Poisson-tessellation where densifying Poisson transmitters does not alter the shape of the association region (the shape of typical Poisson-Voronoi cell is scale invariant), in vehicular networks, the densification of transmitters produces completely new association regions, thin rectangles.     
\end{remark}

\section{Conclusion}
This paper presents  properties of the Poisson line Cox point process seen as a spatial model of vehicular networks. We show that the point process is in general quadratic position. We give the nearest distance distribution and the Laplace functional of the point process, including under the Palm distribution. In particular, we prove that as $ \mu\to\infty $ the typical Cox-Voronoi cell almost surely converges to a segment whose length is given by an Erlang distribution with parameter $ (2,2\lambda_l)$. The results presented in this paper can be used to quantify the performance of new wireless architectures based on  vehicular transmitters located on a random road network. 

\section*{Acknowledgment}
{\small{This work is supported in part by the National Science Foundation under Grant
		No. NSF-CCF-1218338 and an award from the Simons Foundation (\#197982), both
		to the University of Texas at Austin.}}

	 \bibliographystyle{IEEEtran}
	 \bibliography{references}
	 \appendices
	 
	 \section{}\label{A:1}
	 	$ \Phi=\sum_i\phi_i $ where $ \phi_i $ is a PPP of intensity $ \mu $ on the line $ l(r_i,\theta_i)$. Then, shifting by $ t=(v,w)\in\bR^2 $ in the Euclidean plane is associated with the shear  $(r_i,\theta_i)\to(r_i+d\sin(\alpha-\theta_i),\theta_i)\in\mathbf{C} $ with $ d=\sqrt{v^2+w^2} $ and $ \alpha=\tan^{-1}(w/v) $ \cite[Sec. 8.2.2]{chiu2013stochastic}. Then, for all $ B\in\bR^2, $ the Laplace transform of $ \Phi(B+t) $ is 
	 	\begin{align*}
	 		\cL_{\Phi(B+t)}(s)&=\bE\left[\exp\left(-s\int_{B+t}\Phi(\diff x)\right)\right]\\
	 		%&\stackrel{}{=}\bE\left[\prod_{i}\exp\left(-\sum_{X_j\in\cS\circ \phi_i}f(X_j)\right)\right]\\
	 		&=\bE\left[\exp\left(-s\int_{B}S_t\Phi(\diff x)\right)\right]\\
	 		&=\bE\left[\bE\left[\left.\prod_i e^{-s\int_B S_t\phi_{i}(\diff x) }\right|\Psi\right]\right]\\
	 		&=\bE\left[\bE\left[\left.\prod_ie^{-\mu s\cdot \ell(B\cap l(r_i+d\sin(\alpha-\theta_i),\theta_i))} \right|\Psi\right]\right],
	 		%&\equiv\bE\left[\prod_i\exp\left(-\int_B f(x)\Lambda_{\phi_i}(\diff x)\right)\right]=\cL_{\Phi}(f),
	 	\end{align*}
where $ \cS_t $ denotes the shift operator on point process. Since the length $\ell(\cdot)  $ is left invariant by the shear, we have $ \cL_{\Phi(B+t)}(s)\equiv\cL_{\Phi(B)}(s) $. 
Rotation invariance of $ \Phi $ can be proved similarly.

	 \section{}\label{A:2}
	 Let $ B_0(1) $ denotes a ball of radius $ 1 $ centered at $ 0. $ Since the point process is stationary, its intensity is given by 
	 \begin{align*}
	 \bE\left[\Phi(B_0(1))\right]&\ea\bE_{\Psi}\left[\left.\sum_{i\in\bZ}\bE\left[\phi_{l(r_i,\theta_i)}\cap B_0(1)\right]\right|\Psi\right]\\
	 &\eb\bE_{\Psi}\left[\sum_{i\in\bZ}\mu\cdot \text{length}\left({l(r_i,\theta_i)}\cap B_0(1)\right)\right]\\
	 &\ec\bE_{\Psi}\left[\sum_{i\in\bZ}\mu\cdot 2\sqrt{1^2-r_i^2}\right]\\
	 %&\stackrel{(d)}{=}{\mu}\int_{C^\star}\nu_1(l(r,\theta)\cap B_0(1))\Lambda_l(\diff r \diff \theta)\\
	 &\stackrel{(d)}{=}{\mu}\int_{-1}^{1}\int_{0}^{\pi}2\sqrt{1-r^2} \frac{\lambda_l}{\pi}\diff r\diff \theta=\mu\lambda_l \pi,
	 \end{align*}
	 where we have (a) from conditioning on $ \Psi $, (b) from the fact that the number of the PPP on a line is a  Poisson random variable with mean  $ \mu $ times the length of interval, (c) from the length of the arc that the line $ l(r_i,\theta_i) $ creates, and (d) from Campbell's mean value formula \cite{chiu2013stochastic}. %As a result, we have density $ \lambda_l\mu. $

	 \section{}\label{A:3}
	 Let $ A $ denote the event that four points lie in the same circle, namely $\{\exists X,Y,Z,W/ \|X-a\|=\|Y-a\|=\|Z-a\|=\|W-a\|, \neq X,Y,Z,W\}$ where $ \neq X,Y,Z,W $ means $ X,Y,Z,W $ are not the same. Due to stationarity, $ A=\{X,Y,Z,W/\|X\|=\|Y\|=\|Z\|=\|W\|, X,Y,Z,W\}. $ \par Let $ \mathbbm{1}_{\cdot,\cdot,\cdot,\cdot} $ denote the combination of four points distributed on four lines. For instance, $ \mathbbm{1}_{1,1,1,1} $ denotes four points are on different lines, $ \mathbbm{1}_{2,2} $ denotes two points are on one line and the two other points are on another line. Since we consider only four points, there exist at most five different combinations. The probability  $\bP( A) $ is  upper bounded by
\begin{align}
\bP(A)&{\leq} \bE\!\!\!\!\!\!\!\!\sum_{X,Y,Z,W\in\Phi}\!\!\!\!\!\!\!\!\mathbbm{1}_A\left(\mathbbm{1}_{1,1,1,1}+\mathbbm{1}_{1,1,2}+\mathbbm{1}_{1,3}+\mathbbm{1}_{2,2}+\mathbbm{1}_{4}\right)\nnb\\
&\stackrel{(a)}{\leq}  \bE\sum\mathbbm{1}_A\mathbbm{1}_{1,1,1,1}+\bE\sum\mathbbm{1}_A\mathbbm{1}_{1,1,2}+\bE\sum\mathbbm{1}_{2,2}\nnb,
\end{align} where   (a) is given by the fact that $ \mathbbm{1}_{A}\mathbbm{1}_{1,3}=\mathbbm{1}_{A}\mathbbm{1}_{4} =\emptyset $. \par In addition, we have 
\begin{align*}
&\bE\left[\sum_{{X,Y,Z,W}\in\Phi} \ind_A\ind_{2,2}\right]\\
%&= \bE\left[\sum_{l_1\neq l_2}\sum_{{\{X,Y\}\in l_1}}\sum_{\{Z,W\}\in l_2}\ind_{A}\right]\\
&\stackrel{(b)}{=} \bE\left[\int_{{(\bC)}^2} \sum_{{\{X,Y\}\in l_1}}\sum_{\{Z,W\}\in l_2}\ind_A \Xi^{(2)}( r \theta, r'\theta')\right]\\
&= \bE\left[\frac{\lambda_l^2}{\pi^2}\int_{{(\bC)}^2}\left(\sum_{{\{X,Y\}\in l_1}}\sum_{\{Z,W\}\in l_2}\ind_A\right)\diff (r \theta)\diff (r' \theta')\right]\\
&\stackrel{(c)}{=} \!\bE\left[\frac{\lambda_l^2}{\pi^2}\int_{{(\bC)}^2}\left( \int_{{(\bR^2)}^2} \!\!\!\!\!\!\!\! \ind_A\diff^2 (xy)  \diff^2 (zw) \right)\!\diff (r \theta)\diff (r' \theta')\right]\!=0,
\end{align*}
where (b) is obtained by using  the reduced second order moment measure of $ \Xi $ \cite{baccelli2009stochastic} and by writing two distinct lines as $ l_1=l_1(r,\theta) $ and $ l_2=l_2(r',\theta') $. Since the Lebesgue measure of the set $ A $ is zero, we have (c). 
\par Using the same technique, we have 
\begin{align*}
\bE\left[\sum_{{X,Y,Z,W}} \ind_A\ind_{1,1,2}\right]=0 \text{ , }\bE\left[\sum_{{X,Y,Z,W}} \ind_A\ind_{1,1,1,1}\right]=0,
%&= \bE\left[\sum_{l_1\neq l_2\neq l_3}\sum_{{\{X\}\in l_1}}\sum_{\{Y\}\in l_2}\sum_{\{Z,W\}\in l_3}\ind_{A}\right]\\
%&= \bE\left[\int\limits_{} \sum_{{X\in l_1}}\sum_{Y\in l_2}\sum_{\{Z,W\}\in l_3}\ind_A \Xi^{(3)} (r \theta,r'\theta',r'' \theta'')\right]=0,\\
%&= \bE\left[\frac{\lambda_l^3}{\pi^3}\int_{\bC\!\times\bC\!\times\bC}\left(\sum_{{X\in l_1}}\sum_{Y\in l_2}\sum_{\{Z,W\}\in l_3}\ind_A\right)\diff(r \theta)\diff(r'\theta')\diff(r'' \theta'')\right]=0,
%\end{align*}
%where we use 
%$$ 
%\sum_{{X\in l_1}}\sum_{Y\in l_2}\sum_{Z,W\in l_3}\ind_A=\int_{{\bR^2}^3}\ind_A \diff x\diff y \diff z=0.
%$$
%\par Finally, we have 
%\begin{align*}
%&\bE\left[\sum_{{X,Y,Z,W}} \ind_A\ind_{1,1,1,1}\right]=0,
%&= \bE\left[\sum_{l_1\neq l_2\neq l_3\neq l_4}\sum_{{X\in l_1}}\sum_{Y\in l_2}\sum_{Z\in l_3}\sum_{W\in l_4}\ind_{A}\right]=0,
%&= \bE\left[\int\int\int\int\sum_{{X\in l_1}}\sum_{Y\in l_2}\sum_{Z\in l_3}\sum_{W\in l_4}\ind_A \Xi^{(4)}(\diff r \diff \theta\diff r' \diff \theta'\diff r'' \diff \theta''\diff r'''\theta''')\right]\\
%&=\bE\left[\int_{\bC\bC\bC\bC}\left(\sum_{{X\in l_1}}\sum_{Y\in l_2}\sum_{Z\in l_3}\sum_{W\in l_4}\ind_A\right)\Xi^{(4)}(r\theta, r' \theta', r'' \theta'',r'''\theta'''\right]=0,
\end{align*}
by using the following facts:
\begin{align*}
&\sum_{{X\in l_1}}\sum_{Y\in l_2}\sum_{Z,W\in l_3}\ind_A=\int_{{\bR^2}^3}\ind_A \diff x\diff y \diff z=0,\\
&\sum_{{X\in l_1}}\sum_{Y\in l_2}\sum_{Z\in l_3}\sum_{W\in l_4}\ind_A=\int_{{(\bR^2)}^4}\ind_A \diff x \diff y \diff z \diff w=0.
\end{align*}

\section{}\label{A:4}
	Let $ \Phi^{(0)},\Phi^{(1)}$, and $\Phi^{(2)} $ denote the vertices, edges, and faces of the Cox-Voronoi tessellation, respectively. Let $ \lambda^{(0)} $, $ \lambda^{(1)} $, and $ \lambda^{(2)} $ denote their densities, respectively. Then, imagine a translation-invariant directed graph $ G $ on $ \Phi^{(0)}+\Phi^{(2)} $ whose edges are directed from $ \Phi^{(0)} $ toward $ \Phi^{(2)} $. Then, the mean out-degree and the mean in-degree of the typical point of $ \Phi^{(0)}+\Phi^{(2)} $ should be the same due to the mass transportation principle (MTP) \cite{baccelli2013elements}. Therefore, we have 
\begin{align}
&\bE G(0^+)=\frac{3 \lambda^{(0)}}{\lambda^{(0)}+\lambda^{(2)}}\stackrel{\text{MTP}}{=}\bE G(0^-)=\frac{C \lambda^{(2)}}{\lambda^{(0)}+\lambda^{(2)}},\label{M:2}
\end{align}
where $ C $ is an unknown constant giving the number of edges of $ G $ directed from vertices to centroids, 2-dimensional facets. 
\par Similarly, suppose an invariant directed graph $ G' $ on $ \Phi^{(1)}+ \Phi^{(2)} $ where edges are directed from  $ \Phi^{(1)} $ to  $ \Phi^{(2)}$. Then, MTP gives 
\begin{align}
\bE G'(0^+)=\frac{2 \lambda^{(1)}}{\lambda^{(1)}+\lambda^{(2)}} \stackrel{\text{MTP}}{=}\bE G'(0^-)=\frac{C \lambda^{(2)}}{\lambda^{(1)}+\lambda^{(2)}}.\label{M:4}
\end{align}
%	where mass transport principle gives Eq. \eqref{M:3} is equals to Eq. \eqref{M:4}. 
Finally, incorporating Eqs. \eqref{M:2}-\eqref{M:4}  with Euler's formula for planar graphs,
$ 	\lambda^{(0)}-\lambda^{(1)}+\lambda^{(2)}=0,
$	completes the proof.	 
	 
	 %	\par Similarly, the distance between two consecutive points on the same line follows the exponential distribution with parameter $ \mu. $ Thus the length between long edges is given by the arithmetic mean of two independent exponential with parameter $ \mu. $ Thus, the distribution is Erlang with parameter $ (2,2\mu). $
	 %	 	If $\mu/\lambda_l  $ is large enough, the typical cell of repository converges to a convex polytope whose edges are given by Voronoi boundary of two nearest Poisson lines and the Voronoi boundaries of two repository process. See Figure \ref{fig:11005stit} for $ \mu/\lambda_l=100 $ for the convergence of typical cell. 
	 %	 	\par Notice the polytope is dissected by the typical line and, if $ \mu/\lambda_l $ tends to infinity, the height of the polytope follows the Erlang distribution with parameter $ (2,2\lambda_l) $ (Lemma \ref{L:TH}). In the same vein, since the repositories follow Poisson point process on the Poisson lines, 
	 %	 	\par From the mean value theorem, the mean area of the derived polytope is $ \frac{1}{\mu\lambda_l} $ for any values of $ \mu $ and $ \lambda_l $.  In addition, as $ \mu/\lambda_l $ tends to infinity, two long edges of the polytope is parallel while the other edges are not. As a result, the typical polytope converges to a trapezoid (rectangle) asymptotically. In the asymptotic region, the number of edges is four almost surely while the mean number of vertices is given by six according to mean value theorem. This alludes that there exist two $ \pi$-vertices on average.  

\end{document}